\documentstyle[11pt]{article}
\begin{document}

\title{\textbf{Power-Law Entropy-Corrected New Agegraphic Dark Energy
in Ho\v{r}ava-Lifshitz Cosmology}}

\author{K. Karami$^{1,2}$\thanks{KKarami@uok.ac.ir} ,
A. Sheykhi$^{3,2}$\thanks{sheykhi@mail.uk.ac.ir} , M.
Jamil$^{4,5}$\thanks{mjamil@camp.nust.edu.pk} ,  R.
Myrzakulov$^{5}$\thanks{rmyrzakulov@csufresno.edu} , S.
Ghaffari$^{6}$,
 A. Abdolmaleki$^{1}$\\\\
$^{1}$\small{Department of Physics, University of Kurdistan,
Pasdaran St., Sanandaj, Iran}\\$^{2}$\small{Research Institute for
Astronomy $\&$ Astrophysics of Maragha (RIAAM), Maragha, Iran}\\
$^{3}$\small{Department of Physics, Shahid Bahonar University, P.O.
Box 76175, Kerman, Iran}\\$^{4}$\small{Center for Advanced
Mathematics and Physics (CAMP), National University} \\\small{of
Sciences and Technology (NUST), Islamabad,
Pakistan}\\$^{5}$\small{Eurasian International Center for
Theoretical Physics},\\\small{Eurasian National University, Astana,
010008, Kazakstan}\\$^{6}$\small{Physics Department and Biruni
Observatory, College of Sciences},\\\small{Shiraz University, Shiraz
71454, Iran}}

\maketitle

\begin{abstract}
We investigate the new agegraphic dark energy (NADE) model with
power-law corrected entropy in the framework of Ho\v{r}ava-Lifshitz
cosmology. For a non-flat universe containing the interacting
power-law entropy-corrected NADE (PLECNADE) with dark matter, we
obtain the differential equation of the evolution of density
parameter as well as the deceleration parameter. To study parametric
behavior, we use an interesting form of state parameter as
function of redshift $\omega_{\Lambda}(z)=\omega_0+\omega_1 z$. We
find that phantom crossing occurs for the state parameter
for a non-zero coupling parameter, thus supporting interacting dark
energy model.
\end{abstract}

\newpage
\section{Introduction}

Observational data of type Ia supernovae (SNeIa) collected by Riess
et al. \cite{riess} in the High-redshift Supernova Search Team and
by Perlmutter et al. \cite{perl} in the Supernova Cosmology Project
Team independently reported that the present observable universe is
undergoing an accelerated expansion phase. The exotic source for
this cosmic acceleration is generally dubbed ``dark energy'' (DE)
which is distinguished from ordinary matter (such as baryons and
radiation), in the sense that it has negative pressure. This
negative pressure leads to the accelerated expansion of the universe
by counteracting the gravitational force. The astrophysical
observations show that about 70\% of the present energy of the
universe is contained in DE. Although the nature and cosmological
origin of DE is still enigmatic at the present, a great variety of
models has been proposed to describe the DE (see e.g., the
reviews \cite{reviews,luca}). Two promising candidates are the
holographic DE (HDE) \cite{Li} and the agegraphic DE (ADE)
\cite{Cai} models which are originated from some considerations of
the features of the quantum theory of gravity.

It is curious to note that the HDE model has its origin
(i.e. definition and derivation) depends on the Bekenstein-Hawking
(BH) entropy-area relationship $S_{\rm BH}=A/4$ of black hole
thermodynamics, where $A$ is the area of the horizon \cite{Coh}.
However, this definition of HDE can be modified (or
corrected) due to the various correction procedures applied to
gravity theories \cite{Banerjee}. For instance, the corrections to
the entropy which appear in dealing with the entanglement
of quantum fields in and out the horizon \cite{Sau}
generate a power-corrected area term in the entropy
expression. The power-law corrected entropy has the form
\cite{pavon1}
\begin{equation}
S=\frac{A}{4}\left[1-K_{\alpha}
A^{1-\frac{\alpha}{2}}\right],\label{ec}
\end{equation}
where $\alpha$ is a dimensionless constant whose value is currently
under debate and determining its unique and precise value requires
separate investigation, and
\begin{equation}
K_\alpha=\frac{\alpha(4\pi)^{\frac{\alpha}{2}-1}}{(4-\alpha)r_c^{2-\alpha}},
\end{equation}
where $r_c$ is the crossover scale. The second term in Eq.
(\ref{ec}) can be regarded as a power-law correction to the
entropy-area law, resulting from entanglement i.e. the wave function
of the field is taken to be a superposition/entanglement of ground
and exited states \cite{Sau}. The entanglement entropy of the ground
state satisfies the BH entropy-area relationship. Only the excited
state contributes to the correction, and more excitations produce
more deviation from the BH entropy-area law \cite{sau1} (also see
\cite{sau2} for a review on the origin of black hole entropy through
entanglement). This lends further credence to entanglement as a
possible source of black hole entropy. The correction term is also
more significant for higher excitations \cite{Sau}. It is important
to note that the correction term falls off rapidly with
increasing $A$. So for large black holes the correction
term falls off rapidly and the BH entropy-area law is recovered,
whereas for the small black holes the correction is significant.

The ADE model is originated from the uncertainty relation of quantum
mechanics together with the gravitational effect in general
relativity (GR). The ADE model assumes that the observed DE comes
from the spacetime and matter field fluctuations in the universe
\cite{Cai}. Following the line of quantum fluctuations of spacetime,
Karolyhazy \cite{kar} proposed that the distance in Minkowski
spacetime cannot be known to a better accuracy than $\delta
t=\varepsilon t_{P}^{2/3}t^{1/3}$, where $\varepsilon$ is a
dimensionless constant of order unity and $t_P$ is the reduced
Planck time. Based on Karolyhazy relation, Maziashvili proposed that
the energy density of metric fluctuations of Minkowski spacetime is
given by \cite{maz}
\begin{equation}\label{2}
\rho_{\Lambda}\sim\frac{1}{t_{P}^{2}t^{2}}\sim\frac{M_{P}^{2}}{t^{2}},
 \end{equation}
 where $M_P$ is the reduced Planck mass $M_P^{-2}=8\pi G$. Since in the original ADE model the
age of the universe is chosen as the length measure, instead of the
horizon distance, the causality problem in the HDE is avoided
\cite{Cai}. The original ADE model had some difficulties. In
particular, it cannot justify the matter-dominated era \cite{Cai}.
This motivated Wei and Cai \cite{Wei2} to propose the new ADE (NADE)
model, while the time scale is chosen to be the conformal time
instead of the age of the universe. The NADE density is given by
\cite{Wei2}
\begin{equation}
\rho_{\Lambda}=\frac{3{n}^2M_P^2}{\eta^2},\label{NADE}
\end{equation}
where 3$n^2$ is the numerical factor and $\eta$ is the conformal
time and defined as
\begin{equation}
\eta=\int\frac{{\rm d}t}{a}=\int_0^a\frac{{\rm
d}a}{Ha^2}.\label{eta}
\end{equation}
The ADE models have been examined and constrained by various
astronomical observations \cite{age2,age3,age,shey1,Jamilnade,
karami2}. Inspired by the power-law corrected entropy relation
(\ref{ec}), and following the derivation of HDE \cite{Gub} and
entropy-corrected HDE (ECHDE) \cite{Wei}, we can easily obtain the
so-called ``power-law entropy-corrected'' NADE (PLECNADE) whose the
scale is chosen to be the conformal time $\eta$. Therefore, we write
down the energy density of PLECNADE as \cite{shey2,esm}
\begin{equation}
\rho_{\Lambda} = \frac{3n^2{M_P^2}}{\eta^2}-\frac{\beta
M_P^2}{\eta^{\alpha}},\label{density-nade}
\end{equation}
where $\beta$ is a dimensional constant whose the precise
value needs to be determined. In this paper,  our aim is to
investigate the PLECNADE model in Ho\v{r}ava-Lifshitz cosmology.

The plane of the paper as follows: In section 2, we give a brief
review of Ho\v{r}ava-Lifshitz cosmology in detailed balance case. In
section 3, we construct a model of interaction between DE and DM. In
section 4, we discuss some cosmological implications of this model.
We obtain the evolution of dimensionless energy density,
deceleration parameter and equation of state parameter of PLECNADE
model. In section 5 we give the conclusion.

\section{Basics of Ho\v{r}ava-Lifshitz Cosmology: Detailed Balance Case}

Recently a power-counting renormalizable UV complete theory of
gravity was proposed by Ho\v{r}ava \cite{horava}. Quantum gravity
models based on an anisotropic scaling of the space and time
dimensions have recently attracted significant attention \cite{att}.
In particular, Ho\v{r}ava-Lifshitz point gravity might be has
desirable features, but in its original incarnation one is forced to
accept a non-zero cosmological constant of the wrong sign to be
compatible with observations \cite{wrong}. At a first look it seems
that this non-relativistic model for quantum gravity has a well
defined IR limit and it reduces to GR. But as it was first indicated
by Mukohyama \cite{muko}, Ho\v{r}ava-Lifshitz theory mimics GR plus
dark matter (DM). This theory has a scale invariant power spectrum
which describes inflation \cite{inflation}. Moreover, some new integrable and
nonintegrable cosmological models of the Ho\v{r}ava-Lifshitz gravity
have been discussed in \cite{ratbay}. Phenomenologically, in Ho\v{r}ava-Lifshitz gravity the radiation energy density decreases proportional to $a^{-6}$ \cite{mukho}. Hence the resultant baryon asymmetry as well as the stochastic gravity waves can be enhanced. Some cosmological solutions to in Ho\v{r}ava-Lifshitz gravity are obtained previously \cite{cai}. Saridakis formulated Horava-Lifshitz cosmology with an additional scalar field and showed that Horava-Lifshitz dark energy naturally presents very interesting behaviors, possessing a varying equation-of-state parameter, exhibiting phantom behavior and allowing for a realization of the phantom divide crossing. In addition, Horava-Lifshitz dark energy guarantees for a bounce at small scale factors and it may trigger the turnaround at large scale factors, leading naturally to cyclic cosmology \cite{saridakis}. The large scale evolution and curvature perturbations in HL gravity are explored in \cite{lss}, while the origin of primordial large-scale magnetic fields in the Horava's non-relativistic gravity have been discussed in \cite{maeda}. The generalized second law of thermodynamics in Horava-Lifshitz cosmology is studied in \cite{jcap}.  For reviews on the scenario
where the cosmological evolution is governed by Ho\v{r}ava-Lifshitz
gravity see \cite{muko,HL}.

Under the detailed balance and the projectability conditions, the
modified Friedmann equations in the framework of Ho\v{r}ava-Lifshitz
(HL) gravity are given by \cite{horava}
\begin{eqnarray}\label{Fr1fluid}
H^2 = \frac{\kappa^2}{6(3\lambda-1)} \rho_{\rm m}
+\frac{\kappa^2}{6(3\lambda-1)}\left[ \frac{3\kappa^2\mu^2
k^2}{8(3\lambda-1)a^4} +\frac{3\kappa^2\mu^2\Lambda
^2}{8(3\lambda-1)}
 \right]-\frac{\kappa^4\mu^2\Lambda  k}{8(3\lambda-1)^2a^2},
\end{eqnarray}
\begin{eqnarray}\label{Fr2fluid}
\dot{H}+\frac{3}{2}H^2 = -\frac{\kappa^2}{4(3\lambda-1)} p_{\rm m}
-\frac{\kappa^2}{4(3\lambda-1)}\left[\frac{\kappa^2\mu^2
k^2}{8(3\lambda-1)a^4} -\frac{3\kappa^2\mu^2\Lambda
^2}{8(3\lambda-1)}
 \right]-\frac{\kappa^4\mu^2\Lambda  k}{16(3\lambda-1)^2a^2},
\end{eqnarray}
where $H=\frac{\dot a}{a}$ is the Hubble parameter, $\lambda$ is a
dimensionless constant and $\Lambda $ is a positive constant which
as usual is related to the cosmological constant in the IR limit.
The parameters $\kappa$ and $\mu$ are constants. Also $k$ denotes
the curvature of space $k=0,1,-1$ for a flat, closed an open
universe, respectively. Furthermore, $\rho_{\rm m}$ and $p_{\rm m}$
are the energy density and pressure of the matter.

Noticing the form of the above Friedmann equations, we can define
the energy density $\rho_\Lambda$ and pressure $p_{\Lambda}$ for DE
as
\begin{equation}\label{rhoDE}
\rho_\Lambda\equiv \frac{3\kappa^2\mu^2 k^2}{8(3\lambda-1)a^4}
+\frac{3\kappa^2\mu^2\Lambda ^2}{8(3\lambda-1)},
\end{equation}
\begin{equation}
\label{pDE} p_{\Lambda}\equiv \frac{\kappa^2\mu^2
k^2}{8(3\lambda-1)a^4} -\frac{3\kappa^2\mu^2\Lambda
^2}{8(3\lambda-1)}.
\end{equation}
The first term on the right hand side proportional to $a^{-4}$ is
effectively the ``dark radiation term'', present in HL cosmology
\cite{Calcagni:2009ar}, while the second term is referred as an
explicit cosmological constant.

Finally in order for these expressions to match with the standard
Friedmann equations we set \cite{Calcagni:2009ar,Mubasher}
\begin{eqnarray}
&&G_{\rm c}=\frac{\kappa^2}{16\pi(3\lambda-1)},\label{simpleconstants0a}\\
&&\frac{\kappa^4\mu^2\Lambda}{8(3\lambda-1)^2}=1,
\label{simpleconstants0}
\end{eqnarray}
where $G_{\rm c}$ is the ``cosmological'' Newton's constant. Note
that in gravitational theories with the violation of Lorentz
invariance (like HL gravity) the ``gravitational'' Newton's constant
$G_{\rm g}$, which is present in the gravitational action,
differs from the ``cosmological'' Newton's constant $G_{\rm
c}$, which is present in the Friedmann equations, unless
Lorentz invariance is restored \cite{Carroll:2004ai}. For the sake
of completeness we write
\begin{eqnarray}
G_{\rm g}=\frac{\kappa^2}{32\pi}\label{Ggrav}.
\end{eqnarray}
Note that in the IR limit ($\lambda=1$), where Lorentz invariance is
restored, $G_{\rm c}$ and $G_{\rm g}$ are the same.

Further we can rewrite the modified Friedmann Eqs.
(\ref{Fr1fluid}) and (\ref{Fr2fluid}) in the usual forms as
\begin{equation}
\label{eqfr} H^2+\frac{k}{a^2} = \frac{8\pi G_{\rm c}}{3}(\rho_{\rm
m}+\rho_\Lambda),
\end{equation}
\begin{equation}
\label{Fr2b} \dot{H}+\frac{3}{2}H^2+\frac{k}{2a^2} = - 4\pi G_{\rm
c}(p_{\rm m}+p_\Lambda).
\end{equation}

\section{Model with Interaction}

Here we would like to investigate the PLECNADE in HL theory. To do
this we consider a spatially non-flat Friedmann-Robertson-Walker
(FRW) universe containing the PLECNADE and DM. Let us define the
dimensionless energy densities as
\begin{equation}
\Omega_{\rm m}=\frac{\rho_{\rm m}}{\rho_{\rm cr}}=\frac{8\pi {\rm
G_{c}}}{3H^2}\rho_{\rm m},~~~~~~\Omega_{\rm
\Lambda}=\frac{\rho_{\Lambda}}{\rho_{\rm cr}}=\frac{8\pi {\rm
G_{c}}}{3H^2}\rho_{\Lambda},~~~~~~\Omega_{k}=-\frac{k}{a^2H^2},
\label{eqomega}
\end{equation}
thus the Friedmann Eq. (\ref{Fr1fluid}) can be rewritten as
\begin{equation}
1-\Omega_{k}=\Omega_{\Lambda}+\Omega_{\rm m}.\label{eq10}
\end{equation}
Taking time derivative of Eq. (\ref{density-nade}) and using
relation $\dot{\eta}=1/a$, we get
\begin{equation}
\dot{\rho}_{\Lambda}=\left(\frac{1}{a\eta}\right)\left[-2\rho_{\Lambda}+\frac{\beta
M_P^2}{\eta^{\alpha}}(\alpha-2)\right].\label{rhodot}
\end{equation}
Also, if we take the time derivative of the second relation in Eq.
(\ref{eqomega}) after using (\ref{rhodot}), as well as relations
$\dot{\eta}=1/a$ and $\dot{\Omega}_{\Lambda}= H
{\Omega}^{\prime}_{\Lambda}$, we obtain the equation of motion for
${\Omega}_{\Lambda}$ as
\begin{eqnarray}
\label{omegaD-eq-motion1} {\Omega^{\prime}_{\Lambda}} =
\left[-2\Omega_{\Lambda}\frac{\dot{H}}{H^2}-\frac{2\Omega_{\Lambda}}{aH\eta}+\frac{G_{\rm
c}}{G_{\rm g}}\frac{\beta(\alpha-2)}{3aH^3\eta^{\alpha+1}}\right].
\end{eqnarray}
Here, prime denotes the derivative with respect to $x=\ln a$. Taking
derivative of $\Omega_{k} = -k/(a^2H^2)$ with respect to $x = \ln
a$, one gets
\begin{equation}
{\Omega^{\prime}_{k}}=-2\Omega_{k}\left(1+\frac{\dot{H}}{H^2}\right).\label{omegak-eq-motion1}
\end{equation}
To be more general we consider an interaction between DM and
PLECNADE. The recent observational evidence provided by the galaxy
clusters supports the interaction between DE and DM
\cite{Bertolami8}. In this case, the energy densities of DE and DM
no longer satisfy independent conservation laws. They obey instead
\begin{equation}
\dot{\rho}_{\Lambda}+3H(1+\omega_{\Lambda})\rho_{\Lambda}=-Q,\label{eqintDE}
\end{equation}
\begin{equation}
\dot{\rho}_{\rm m}+3H\rho_{\rm m}=Q,\label{eqintCDM}
\end{equation}
where $Q=3b^2H\rho_{\Lambda}$ stands for the interaction term with
coupling constant $b^2$. Note that the form of Q is chosen purely
phenomenologically i.e. to obtain certain desirable cosmological
findings including phantom crossing and accelerated expansion. In
literature, one can find numerous forms of $Q(H\rho)$ while we chose
a simpler form sufficient for our purpose. A more general form of
$Q$ was proposed in \cite{rashid}. Also the three
interacting fluids has been studied before to investigate the
triple coincidence problem \cite{triple}. Differentiating
the Friedmann Eq. (\ref{Fr1fluid}) with respect to time and
using Eqs. (\ref{eqomega}), (\ref{eq10}), (\ref{rhodot}),
(\ref{eqintDE}) and (\ref{eqintCDM}) we find
\begin{eqnarray}
\frac{\dot{H}}{H^2} =\frac{1}{2}\Big[\Omega_k-3(1 -
\Omega_{\Lambda})+3b^2\Omega_{\Lambda}\Big]
-\frac{\Omega_{\Lambda}}{aH\eta}+\frac{G_{\rm c}}{G_{\rm
g}}\frac{\beta(\alpha-2)}{6aH^3\eta^{\alpha+1}}.\label{H-dot-to-H2-interact}
\end{eqnarray}
Inserting this result into Eq. (\ref{omegaD-eq-motion1})
one gets
\begin{eqnarray}
{\Omega^{\prime}_{\Lambda}}= \Omega_{\Lambda} \left[3(1 -
\Omega_{\Lambda})-3b^2{\Omega}_{\Lambda} -\Omega_k\right]+(1 -
\Omega_{\Lambda})\left[\frac{-2\Omega_{\Lambda}}{aH\eta}+\frac{G_{\rm
c}}{G_{\rm
g}}\frac{\beta(\alpha-2)}{3aH^3\eta^{\alpha+1}}\right].\label{omegaD-eq-motion3}
\end{eqnarray}
Combining Eq. (\ref{H-dot-to-H2-interact}) with
(\ref{omegak-eq-motion1}) we have
\begin{equation}
{\Omega^{\prime}_{k}}=\Omega_{k}\left[(1
-\Omega_k)-3\Omega_{\Lambda}-3b^2\Omega_{\Lambda}
+\frac{2\Omega_{\Lambda}}{aH\eta}-\frac{G_{\rm c}}{G_{\rm
g}}\frac{\beta(\alpha-2)}{3aH^3\eta^{\alpha+1}}\right].\label{omega-prime-k-interact}
\end{equation}
Adding Eqs. (\ref{omegaD-eq-motion3}) and
(\ref{omega-prime-k-interact}) yields
\begin{eqnarray}
{\Omega^{\prime}_{\Lambda}}+{\Omega^{\prime}_{k}}=(1-\Omega_{k}-\Omega_{\Lambda})\left[\Omega_{k}+3\Omega_{\Lambda}
-\frac{3b^2\Omega_{\Lambda}(\Omega_{k}+\Omega_{\Lambda})}{(1-\Omega_{k}-\Omega_{\Lambda})}+\frac{G_{\rm
c}}{G_{\rm
g}}\frac{\beta(\alpha-2)}{3aH^3\eta^{\alpha+1}}-\frac{2\Omega_{\Lambda}}{aH\eta}\right].\label{omegalambdakpri}
\label{add-omega-prime-interact}
\end{eqnarray}
For completeness we give the deceleration parameter which is defined
as
\begin{equation}
q=-\left(1+\frac{\dot{H}}{H^2}\right).\label{q1}
\end{equation}
After combining Eq. (\ref{H-dot-to-H2-interact}) with (\ref{q1}) we
get
\begin{eqnarray}
q = \frac{1}{2}[1-\Omega_{k}-3(1+b^2)\Omega_{\Lambda}]+
\frac{\Omega_{\Lambda}}{aH\eta}-\frac{G_{\rm c}}{G_{\rm
g}}\frac{\beta(\alpha-2)}{3aH^3\eta^{\alpha+1}}.
\end{eqnarray}
Using definitions (\ref{eqomega}) as well as Eq. (\ref{eq10}) we
have
\begin{equation}
\rho_{\Lambda}=\frac{\rho_{\rm m}}{\Omega_{\rm
m}}\Omega_{\Lambda}=\frac{\rho_{\rm
m}}{(1-\Omega_{k}-\Omega_{\Lambda})}\Omega_{\Lambda},\label{rholambdaint}
\end{equation}
which from it we can obtain
\begin{equation}
\frac{\rm d{\ln{\rho_{\Lambda}}}}{{\rm d}
\ln{a}}=\frac{\rho^{\prime}_{\rm m}}{\rho_{\rm
m}}-\frac{\Omega^{\prime}_{\rm m}}{\Omega_{\rm
m}}+\frac{\Omega^{\prime}_{\Lambda}}{\Omega_{\Lambda}}.\label{drholambda}
\end{equation}

\section{Cosmological Implications}

In this section, we study some cosmological consequences of a
phenomenologically time-dependent parameterization for the PLECNADE
equation of state as
\begin{equation}
\omega_{\Lambda}(z)=\omega_0+\omega_1 z.\label{wpar}
\end{equation}
It was shown in \cite{barboza} that this parameterization
allows to divide the parametric plane $(\omega_0,\omega_1)$ in
defined regions associated to distinct classes of DE models that can
be confirmed or excluded from a confrontation with current
observational data.

After using Eq. (\ref{eqintDE}), the evolution of the DE density is
obtained as \cite{reviews,hunt}
\begin{equation}
\frac{\rho_{\Lambda}}{\rho_{\Lambda_{0}}}=a^{-3(1+\omega_{0}-\omega_{1}+b^2)}e^{3\omega_{1}z}.\label{eqCDEint}
\end{equation}
The Taylor expansion of the DE density around $a_0 = 1$ at the
present time yields
\begin{equation}
\ln{\rho_{\Lambda}}=\ln{\rho_{\Lambda_{0}}}+\frac{\rm
d{\ln{\rho_{\Lambda}}}}{{\rm d}
\ln{a}}\Big{|}_0\ln{a}+\frac{1}{2}\frac{\rm d^2
\ln{\rho_{\Lambda}}}{{\rm
d}({\ln{a}})^2}\Big{|}_0(\ln{a})^2+\cdots.\label{taylor expand}
\end{equation}
Using the fact that for small redshifts, $\ln a = -\ln(1 + z) \simeq
-z + \frac{z^2}{2}$, Eqs. (\ref{eqCDEint}) and (\ref{taylor
expand}), respectively, reduce to
\begin{equation}
\frac{\ln{(\rho_{\Lambda}/\rho_{\Lambda_{0}})}}{\ln{a}}=-3(1+\omega_{0}+b^2)-\frac{3}{2}\omega_{1}z\label{eqCDE1int},
\end{equation}
\begin{equation}
\frac{\ln{(\rho_{\Lambda}/\rho_{\Lambda_{0}})}}{\ln{a}}=\frac{\rm
d{\ln{\rho_{\Lambda}}}}{{\rm d}
\ln{a}}\Big{|}_0-\frac{1}{2}\frac{\rm d^2 \ln{\rho_{\Lambda}}}{{\rm
d}({\ln{a}})^2}\Big{|}_0z\label{eqCDE2}.
\end{equation}
Comparing Eq. (\ref{eqCDE1int}) with (\ref{eqCDE2}), we find that
these two equations are consistent provided we have
\begin{equation}
\omega_{0}=-\frac{1}{3}\frac{\rm d{\ln{\rho_{\Lambda}}}}{{\rm d}
\ln{a}}\Big{|}_0-1-b^2,\label{w0int}
\end{equation}
\begin{equation}
\omega_{1}=\frac{1}{3}\frac{\rm d^2 \ln{\rho_{\Lambda}}}{{\rm d}
({\ln{a}})^2}\Big{|}_0.\label{w1int}
\end{equation}
Inserting Eq. (\ref{drholambda}) in (\ref{w0int}) and (\ref{w1int}),
after using (\ref{eqintCDM}), yields
\begin{equation}
\omega_{0}=-\frac{1}{3}\left[\frac{\Omega^{\prime}_{\Lambda}}{\Omega_{\Lambda}}
+\frac{\Omega^{\prime}_{\Lambda}+\Omega^{\prime}_{k}}{(1-\Omega_{k}-\Omega_{\Lambda})}\right]_0
-b^2\left(\frac{1-\Omega_{k}}{1-\Omega_{k}-\Omega_{\Lambda}}\right)_0,
\label{eqomega0-interact}
\end{equation}
\begin{eqnarray}
\omega_{1}=\frac{1}{3}\left[\frac{3b^2\Omega^{\prime}_{\Lambda}}{(1-\Omega_{k}-\Omega_{\Lambda})}
+\frac{3b^2\Omega_{\Lambda}(\Omega^{\prime}_{\Lambda}+\Omega^{\prime}_{k})}{(1-\Omega_{k}-\Omega_{\Lambda})^2}
+\frac{{\Omega}^{\prime\prime}_{\Lambda}}{\Omega_{\Lambda}}
-\frac{{\Omega}^{\prime2}_{\Lambda}}{\Omega_{\Lambda}^2}
\right.~~~~~~~~\nonumber\\\left.+\frac{\Omega^{\prime\prime}_{\Lambda}+{\Omega}^{\prime\prime}_{k}}{(1-\Omega_{k}-\Omega_{\Lambda})}
+\frac{(\Omega^{\prime}_{\Lambda}+\Omega^{\prime}_{k})^2}{(1-\Omega_{k}-\Omega_{\Lambda})^2}\right]_0.\label{w1-2int}
\end{eqnarray}
Substituting Eqs. (\ref{omegaD-eq-motion3}) and
(\ref{add-omega-prime-interact}) into (\ref{eqomega0-interact}) we
reach
\begin{eqnarray}
\label{state-parameter-2-interact}
\omega_{0}=\frac{-1}{3\Omega_{\Lambda_0}}\left(\frac{G_{\rm
c}}{G_{\rm
g}}\frac{\beta(\alpha-2)}{3H_0^3\eta_0^{\alpha+1}}-\frac{2\Omega_{\Lambda_0}}{H_0\eta_0}\right)-b^2-1.
\end{eqnarray}
Taking derivative of Eqs. (\ref{omegaD-eq-motion3}) and
(\ref{omega-prime-k-interact}) with respect to $x = \ln a$ and using
(\ref{state-parameter-2-interact}), one gets
\begin{eqnarray}
\Omega^{\prime\prime}_{\Lambda}&=&-(\Omega_{\Lambda}\Omega^{\prime}_{k}+\Omega^{\prime}_{\Lambda}\Omega_{k})
-3\Omega_{\Lambda}(\Omega_{\Lambda}-\Omega^{\prime}_{\Lambda}-1)(\omega_{0}+b^2+1)
+(1-\Omega_{\Lambda})A
\nonumber\\&&-6\Omega^{\prime}_{\Lambda}\Omega_{\Lambda}(b^2+1)+3\Omega^{\prime}_{\Lambda},\label{omegalambdapri-interact}
\end{eqnarray}
and
\begin{eqnarray}
\Omega^{\prime\prime}_{k}=\Omega^{\prime}_{k}(1-2\Omega_{k})
-3\Omega_{\Lambda}(\Omega_{k}-\Omega^{\prime}_{k})(\omega_{0}+b^2+1)
-A\Omega_{k} -3(\Omega^{\prime}_{k}\Omega_{\Lambda}+\Omega_{
k}\Omega^{\prime}_{\Lambda})(b^2+1),\label{omegakpri-interact}
\end{eqnarray}
where $A$ is given by
\begin{eqnarray}
A=\frac{2\Omega_{\Lambda}}{aH\eta}\left(\frac{\dot{H}}{H^2}+\frac{\dot{\eta}}{H\eta}\right)
-\frac{2\Omega^{\prime}_{\Lambda}}{aH\eta}-\frac{G_{\rm c}}{G_{\rm
g}}\frac{\beta(\alpha-2)}{3aH^3\eta^{\alpha+1}}\left(\frac{3\dot{H}}{H^2}+\frac{(\alpha+1)\dot{\eta}}{H\eta}\right).\label{A}
\end{eqnarray}
The above expression for $A$ can also be rewritten as
\begin{eqnarray}
A&=&\frac{2}{aH\eta}\Big[\Omega_{
k}\Omega_{\Lambda}-3\Omega_{\Lambda}(1-\Omega_{\Lambda}-b^2\Omega_{\Lambda})\Big]
+3\Omega_{\Lambda}(\omega_{0}+b^2+1)\left[\frac{1}{aH\eta}-2\Omega_{\Lambda}-q-1\right]
\nonumber
\\&&-\frac{G_{\rm c}}{G_{\rm
g}}\frac{\beta(\alpha-2)}{3aH^3\eta^{\alpha+1}}\left[\frac{\alpha}{aH\eta}-2(1+q)\right],\label{A2}
\end{eqnarray}
where we have used Eqs. (\ref{H-dot-to-H2-interact}),
(\ref{omegaD-eq-motion3}) as well as the relation $\dot{\eta}=1/a$.
Adding Eqs. (\ref{omegalambdapri-interact}) and
(\ref{omegakpri-interact}) gives
\begin{eqnarray}
\Omega^{\prime\prime}_{\Lambda}+\Omega^{\prime\prime}_{k}&=&(1-\Omega_{\Lambda}-\Omega_{k})\Big[\Omega^{\prime}_{
k}+3\Omega_{\Lambda}(1+\omega_{0}+b^2)+A\Big]+(3\Omega_{\Lambda}\omega_{0}-\Omega_{k})(\Omega^{\prime}_{
k}+\Omega^{\prime}_{\Lambda})
\nonumber\\&&+3\Omega^{\prime}_{\Lambda}\Big[1-(\Omega_{k}+\Omega_{\Lambda})(1+b^2)\Big].\label{omegalambdakppri-interact}
\end{eqnarray}
Finally, by combining Eqs. (\ref{omegaD-eq-motion3}),
(\ref{add-omega-prime-interact}), (\ref{omegalambdapri-interact}),
(\ref{A2}) and (\ref{omegalambdakppri-interact}) with
(\ref{w1-2int}) we find
\begin{equation}
\omega_{1}=(1+\omega_{0}+b^2)\Big[3\omega_{0}(\Omega_{\Lambda_0}-1)-\Omega_{
k_0}-3b^2+1\Big] +\frac{A_0}{3\Omega_{\Lambda_0}},
\end{equation}
which more explicitly can be written as
\begin{eqnarray}
\omega_{1}&=&(1+\omega_{0}+b^2)\Big[\Omega_{\Lambda_0}(3\omega_{0}-2)-3(\omega_0+b^2)-\Omega_{
k_0}-q_0\Big]\nonumber
\\&&+\frac{1}{H_0\eta_0}\left[\omega_0+(1+b^2)(1+2\Omega_{\Lambda_0})+2\left(\frac{\Omega_{k_0}}{3}-1\right)\right]\nonumber
\\&&-\frac{G_{\rm c}}{G_{\rm
g}}\frac{\beta(\alpha-2)}{9H_0^3\eta_0^{\alpha+1}\Omega_{\Lambda_0}}\left[\frac{\alpha}{H_0\eta_0}-2(1+q_0)\right].\label{EoS1}
\end{eqnarray}
Therefore, with $\omega_0$ and $\omega_1$ at hand we can easily
write down the explicit expression for $\omega_{\Lambda}(z)$ in Eq.
(\ref{wpar}) in terms of model parameters such as $\Omega_\Lambda$,
$\Omega_{k}$, the running parameter $\lambda$ of HL gravity, the
parameter $n$ of PLECNADE, the interaction coupling $b^2$, and the
correction coefficients $\alpha$ and $\beta$. From Fig. 1 we notice
that in the absence of interaction $b^2=0$, $\omega_0=-0.99$ showing
quintessence state. By introducing interaction term, the state
parameter evolves to a phantom state and gradually goes to more
super-phantom state. Figure 1 also shows that the phantom
crossing for $\omega_0$ happens for $b^2=0.01$ which is compatible
with the observation \cite{Komatsu}. This is also in agreement with
the result obtained by \cite{Karami3}. Note that phantom crossing has sound empirical support: analysis of Gold SNe and other observational datasets suggests that $\omega(z)\leq-1$, for $0\leq z\leq0.5$ \cite{obser}. From Fig. 2, we notice that
in the absence of interaction, the first order correction $\omega_1$
to state parameter behaves like quintessence but when interaction is
introduced, the parameter $\omega_1$ evolves towards $-1$,
cosmological constant. Here it is probable that $\omega_1$ can cross
the cosmological constant boundary if $b^2>0.20$. For reader's clarity, we did not use any initial conditions for plotting both figures since Eq. (\ref{state-parameter-2-interact}) and Eq. (\ref{EoS1}) are not differential equations. However the behavior of curves in figures is strongly sensitive to the values of free parameters. 

\section{Conclusions}

It has been shown that the origin of black hole entropy may lie in
the entanglement of quantum fields between inside and outside of the
horizon \cite{Sau}. Since the modes of gravitational fluctuations in
a black hole background behave as scalar fields, one is able to
compute the entanglement entropy of such a field, by tracing over
its degrees of freedom inside a sphere. In this way the authors of
\cite{Sau} showed that the black hole entropy is proportional to the
area of the sphere when the field is in its ground state, but a
correction term proportional to a fractional power of area results
when the field is in a superposition of ground and excited states.
For large horizon areas, these corrections are relatively small and
the BH entropy-area law is recovered.

Here, we investigated the PLECNADE scenario in the framework of HL
gravity. We considered an arbitrary spatial local curvature for the
background geometry and allowed for an interaction between the
PLECNADE and DM. We obtained the deceleration parameter as well as
the differential equation which determines the evolution of the
PLECNADE density parameter. Using a low redshift expansion of the
EoS parameter of PLECNADE as
 $\omega_{\Lambda}(z)=\omega_0 +\omega_1
z$, we calculated $\omega_0$ and $\omega_1$ as functions of the
PLECNADE and curvature density parameters, $\Omega_\Lambda$ and
$\Omega_{k}$ respectively, of the running parameter $\lambda$ of HL
gravity, of the parameter $n$ of PLECNADE, of the interaction
coupling $b^2$, and of the coefficients of correction terms $\alpha$
and $\beta$. It is quite interesting to note that phantom crossing
for $\omega_0$ happens for $b^2=0.01$ i.e. a small but non-zero
interaction parameter.  In literature, there are two well-studied ways for phantom crossing: via modified gravities including scalar-tensor or Gauss-Bonnet braneworld models \cite{gauss} or by introducing an interaction between fluid dark energy and dark matter. In the later models, its a generic feature of interacting dark energy models to have phantom crossing, for instance, using different forms of dark energy including Chaplygin gas \cite{sad}, quintessence \cite{sad1}, new-agegraphic and holographic dark energy \cite{Jamilnade}. In \cite{nadebd}, the present authors studied the PLECNADE interacting with matter in Brans-Dicke gravity and obtained the state parameter behaving to cross the phantom divide for small values of coupling parameter $b$.

\subsection*{Acknowledgements}

The works of K. Karami and A. Sheykhi have been supported
financially by Research Institute for Astronomy and Astrophysics of
Maragha (RIAAM) under research project No. 1/2340.  M. Jamil would like to thank the warm hospitality of Eurasian National University, Astana, Kazakhstan where part of this work completed. Authors would thank the anonymous referees for their constructive criticism on this paper.

\clearpage
\begin{figure}
\includegraphics{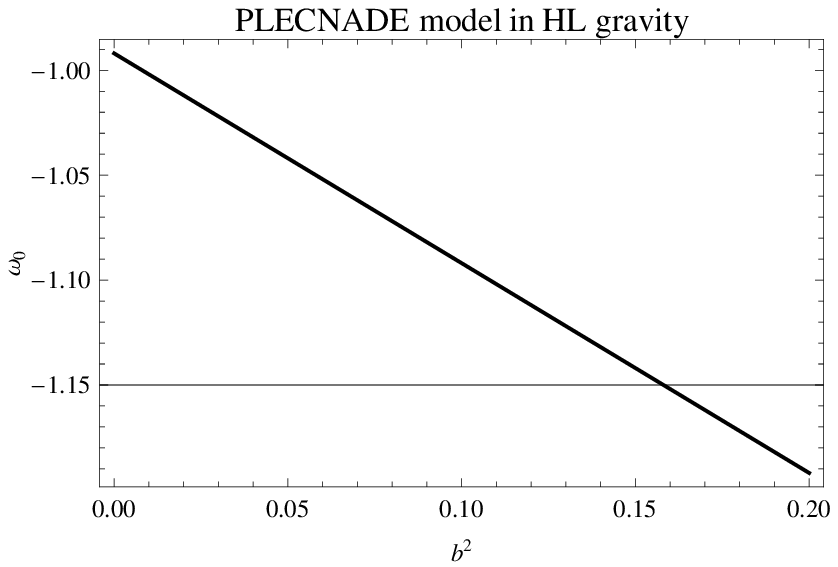}
      \vspace{5cm}
\caption[]{The EoS parameter $\omega_0$ of the PLECNADE in HL
gravity, Eq. (\ref{state-parameter-2-interact}), versus the
interacting coupling parameter $b^2$. Auxiliary parameters are:
$n=2.716$ \cite{age2}, $\alpha=-7.5$, $\beta=-14.8$, $\eta_{0}=1.1$
\cite{Abdolmaleki}, $\Omega_{\Lambda_0}=0.728$,
$\Omega_{k_0}=-0.013$ \cite{Komatsu}, $\lambda=1.02$ \cite{Dutta},
$G_{\rm c}/G_{\rm g}=2/(3\lambda-1)=0.97$, $H_0=74.2~{\rm
Km~S^{-1}~Mpc^{-1}}$ \cite{Riess09} and $M_P^{-2}=8\pi G_{\rm
g}=1$.}
         \label{w0-b2}
   \end{figure}
 \begin{figure}
\includegraphics{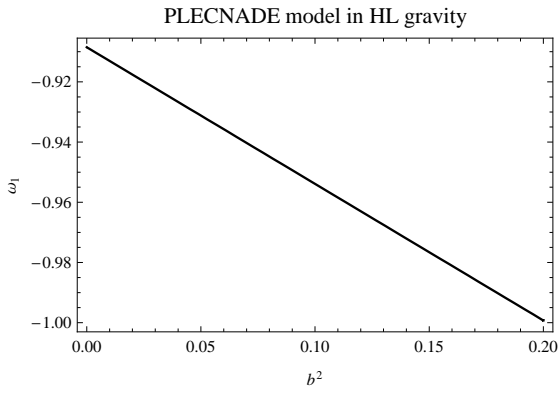}
      \vspace{5cm}
\caption[]{The EoS parameter $\omega_1$ of the PLECNADE in HL
gravity, Eq. (\ref{EoS1}), versus the interacting coupling parameter
$b^2$. Auxiliary parameters as in Fig. \ref{w0-b2}.}
         \label{w1-b2}
   \end{figure}
\end{document}